\def\tr{\mathop{\rm Tr}}

\def\barkappa{{\overline\kappa}}
\def\vect#1{\mathbf{#1}}
\def\V{\hat{\cal V}}

\makeatletter
\def\Hy@safe@activestrue{}
 \makeatother

\documentclass[aps,prl,superscriptaddress,twocolumn,amssymb,%
shortbibliography,%
  amsmath,showpacs]{revtex4-1}
\usepackage{amssymb}
\usepackage{graphicx}
\usepackage[hypertex]{hyperref}
\begin{document}
\title{Fluctuation-induced forces between inclusions in a fluid membrane under
  tension} 
\author{Hsiang-Ku Lin} 
\author{Roya Zandi} 
\author{Umar Mohideen}
\author{Leonid P. Pryadko}

\affiliation{Department of Physics \& Astronomy, University of
  California, Riverside, California 92521, USA}

\date{\today}
%\date{\VERSION}
\begin{abstract}
We develop an exact method to calculate thermal Casimir forces between
inclusions of arbitrary shapes and separation, embedded in a fluid membrane
whose fluctuations are governed by the combined action of surface tension,
bending modulus, and Gaussian rigidity. Each object's shape and
mechanical properties enter only through a characteristic matrix, a static
analog of the scattering matrix. We calculate the Casimir interaction
between two elastic disks embedded in a membrane. In particular, we find
that at short separations the interaction is strong and independent of 
surface tension.
\end{abstract}
%.  The model is applicable to the most general case in which membranes 
% , with the embedded objects of arbitrary shapes and separation
\pacs{%
87.15.kt,% Dynamics and fluctuations
87.16.dj,% Protein-membrane interactions
34.20.-b}% Interatomic and intermolecular potentials and forces, potential 
%%% energy surfaces for collisions

\advance\textheight by 0.1in
%\advance\textheight by -3.7in % notebook 
%\advance\textheight by -1.6in % work 
\renewcommand{\dbltopfraction}{0.95}
\renewcommand{\topfraction}{0.9}
\renewcommand{\textfraction}{0.07}

\maketitle

Biological membranes are formed by bilayers of amphiphilic lipid molecules
with embedded macromolecules such as proteins~\cite{Alberts-book}.  While
inclusion can dramatically modify the flexibility, surface tension, and
spontaneous curvature of biomembranes \cite{Reynwar-2007}, the interactions
between embedded proteins mediated by the membrane, in turn, depend on its
mechanical properties \cite{Dan-Pincus-Safran-1993,%
  Goulian-Bruinsma-Pincus-1993,%
  *Goulian-Bruinsma-Pincus-1993B,%
  Golestanian-Goulian-Kardar-EPL-1996,%
  *Golestanian-Goulian-Kardar-PRE-1996,*Park-Lubensky-1996,%
  Weikl-Kozlov-Helfrich-1998,Weikl-EPL-2001}.  Generally, these interactions
depend on the separation between the inclusions and can be categorized as
elastic and fluctuation-induced~\cite{Goulian-Bruinsma-Pincus-1993,%
  *Goulian-Bruinsma-Pincus-1993B}.  The former is due to static elastic
deformations induced in the membrane, while the latter, the Casimir-like
force, is due to the modified entropy of the membrane fluctuations.

Despite being the subject of intense research for almost two
decades\cite{Goulian-Bruinsma-Pincus-1993,*Goulian-Bruinsma-Pincus-1993B,%
  Golestanian-Goulian-Kardar-EPL-1996,%
  *Golestanian-Goulian-Kardar-PRE-1996,*Park-Lubensky-1996,%
  Noruzifar-Oettel-2009,%
  Yolcu-Rothstein-Deserno-2010}, fluctuation-induced forces are just beginning
to be understood.  Previous theoretical research only considered fluctuations
in the separate presence of either surface tension or the bending rigidities,
but not the two combined\cite{Bartolo-2002}, as occurs in reality.  While
surface tension is negligible for free-floating
membranes\cite{Goulian-Bruinsma-Pincus-1993,*Goulian-Bruinsma-Pincus-1993B},
it is finite for a membrane enclosing a cell with excess osmotic pressure, or
for membranes under tension in general\cite{Rawicz-2000}.  In such
cases, the surface tension $\sigma_0$, as well as the bending ($\kappa_0$) and
Gaussian ($\barkappa_0\equiv \mu_0\kappa_0$) rigidities of the membrane are
all non-zero.
For such membranes, the static elastic interaction between inclusions is
exponentially cut-off at distances larger than the characteristic
length\cite{Weikl-Kozlov-Helfrich-1998,Evans-Turner-Sens-2003}
  \begin{equation}
    \ell_0\equiv \alpha_0^{-1}=(\kappa_0/\sigma_0)^{1/2}.
\label{eq:crossover-length}
\end{equation}

In this letter, we present an exact method that allows calculations of
membrane-mediated fluctuation-induced forces between any number of arbitrarily
placed elastic inclusions of any shape for finite $\ell_0$.  It is based on
the technique previously employed to calculate the Casimir forces in
antiferromagnets\cite{Pryadko-Kivelson-Hone-98}.  It can also be regarded as a
version of the scattering-matrix approach\cite{Rahi-2009}, generalized to
entropic forces for Hamiltonians with quartic in derivatives terms
representing bending energy.  We find that the Casimir energy between objects
embedded in membrane could be expressed in terms of the response of individual
objects to the fluctuating field.  An object's shape and material properties
enter only through the coefficient matrix representing the object's response,
which is a static analog of the scattering matrix.
%%%  object's 
%%% fluctuating fields' scattering amplitudes from the individual objects. The
%%% objects' geometry and elastic properties enter only through the
%%% corresponding 
%%% scattering matrices.
%%% While the calculation of the scattering amplitudes is not trivial,
%%% they can be derived for a number of geometries.

\begin{figure}[!t]
  \includegraphics[width=0.9\columnwidth]{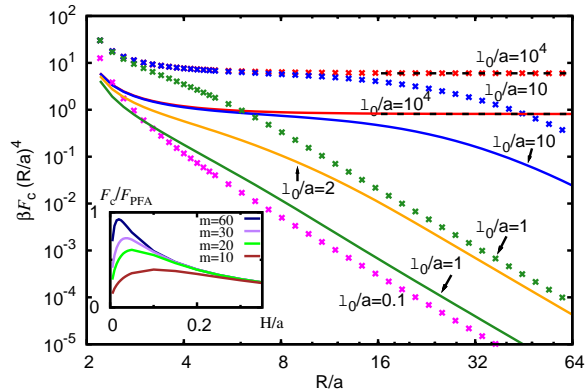}
  \caption{(color online) The Casimir free-energy~(\ref{eq:twodisk-full-result})
    scaled with fourth power of the distance as a function of $R/a$ for
    $\kappa/\kappa_0=10^4$ (symbols) and $\kappa/\kappa_0=10^{-1}$ (solid
    lines), with $\alpha_0 a\equiv a/\ell_0$ as indicated.  We set
    $\barkappa_0=-\kappa_0$, $\barkappa=-\kappa$.  Dashed black lines indicate
    asymptotic large-$R$ dependences evaluated with
    Eq.~(\ref{eq:large-ell-limit}).  Plots are horizontal
    ($\Delta\mathcal{F}\propto R^{-4}$) only for $R\lesssim \ell_0\equiv
    \alpha_0^{-1}$; for larger $R/\ell_0$ the Casimir energy decays as
    $\Delta\mathcal{F}\propto R^{-8}$, see Eq.~(\ref{eq:scaling}). Inset: The
    Casimir energy of two disks in the regime dominated by the bending energy
    ($\ell_0/a=10^4$, $\kappa/\kappa_0=10^4$) divided by the PFA estimate
    (\ref{eq:pfa}). The quantity $m$ indicates the multipole order of
    truncation.}
  \label{fig:scaling}
\end{figure} 

In particular, we compute fluctuation-induced interactions between two
elastic disks. We specifically examine the unexplored parameter range
where $\ell_0$ is between the disk radius $a$ and the center-to-center
separation distance $R$, $a\lesssim \ell_0\lesssim R$, so that both
the surface tension and bending energies are relevant
[Fig.~\ref{fig:scaling}].  In the limit of very short separations, we
find a very good match between our results and those obtained through
Derjaguin\cite{Derjaguin-1956,*Lehle-Oettel-2007} or proximity force
approximation (PFA) [see the inset in Fig.~\ref{fig:scaling}].  In the
large separation limit, $R\gg a$, we derive explicit asymptotics for
the Casimir interaction, and check them against the
known results\cite{Goulian-Bruinsma-Pincus-1993,%
  *Goulian-Bruinsma-Pincus-1993B,%
  Golestanian-Goulian-Kardar-EPL-1996,*Golestanian-Goulian-Kardar-PRE-1996,%
  *Park-Lubensky-1996,Noruzifar-Oettel-2009,Yolcu-Rothstein-Deserno-2010} for cases dominated by either surface tension 
or bending rigidities.

%%% We report the result of our calculations in Fig.~\ref{fig:scaling}, where
%%% we plot the distance-dependence of the parameter $A$ which appears as an
%%% coefficient in the scaling relation of Eq. (\ref{eq:scaling}) [see below]
%%% for the Casimir energy.

For a non-zero surface tension, we find that the Casimir energy becomes
strongly suppressed at distances $R$ larger than the characteristic length
$\ell_0$, Eq.~(\ref{eq:crossover-length}).  Note that the Casimir energy
retains the 
power-law asymptotic form
\begin{equation}
  \label{eq:scaling}
  {\beta\, \mathcal{F}_{\rm C}}=-A \, {a^n/ R^n},\quad
  \beta\equiv 1/k_BT, \quad R\gg a,
\end{equation}
with the exponent increasing from $n=4$ at $R\lesssim \ell_0$
(bending-energy-dominated regime) to $n=8$ at $R\agt\ell_0$ (tension-dominated regime).  This is illustrated in Fig.~\ref{fig:scaling} for
two inclusion stiffnesses, with a number of different surface tensions.
Depending on the parameters, the Casimir free energy scaled with the fourth
power of the distance has either constant or $\propto 1/R^4$ asymptotics.

As seen in the figure, all curves with the same ratio of $\kappa/\kappa_0$
merge at short separations, while at larger separations the value of
$l_0/a\propto \sigma_0^{-1/2}$ plays an important role.  We also find a
surprising effect corresponding to the difference between the Gaussian
rigidities of the membrane and the inclusions.  In the absence of line
tension\cite{Lehle-Oettel-2008} for the inclusion boundaries
[$\sigma=\sigma_0$ in Eq.~(\ref{eq:Helfrich})], the Casimir force at large
distances becomes exponentially small when the Gaussian rigidities of the
inclusions and the membrane coincide, $\barkappa=\barkappa_0$, regardless of
$\kappa$.

We write the energy of a given configuration as a combination of that of the
inclusions (thin isotropic elastic solids characterized by in-plane Lam\'e
coefficients and bending rigidities) and the fluid membrane outside the
inclusions (defined by the sum of the surface tension and bending energies).
The energy is expanded to quadratic order in the displacements with respect to
the equilibrium membrane configuration assumed to be planar in the $z=0$
plane. The result for the membrane with inclusions has the standard Helfrich
form \cite{Canham-1970,*Helfrich-1973},
\begin{equation}
U\equiv\int_{\mathcal{A}} d^2 \vect{r}\; \frac{\sigma}{2} (\nabla
u)^2+
%(\kappa+\barkappa)(\nabla^2u)^2-
%\barkappa(\partial_i\partial_ju)^2, 
\frac{\kappa}{2}(\nabla^2u)^2+\barkappa
[u_{xx}''u_{yy}''-(u_{xy}'')^2],
%%%[\partial_x^2 u \partial_y^2 u -(\partial_x \partial_y u )^2],
\label{eq:Helfrich}
\end{equation}
where primes ($'$) denote the partial derivatives with respect to $x$ or $y$
as indicated, and the integration is done over the total projected area ${\cal
  A}$.  The coefficients $\kappa$ and $\barkappa$ in Eq.~(\ref{eq:Helfrich})
are position-dependent, e.g., $\kappa\equiv \kappa(\mathbf{r})$,
$\mathbf{r}\equiv (x,y)$.  Thermodynamical stability requires that
$\sigma,\kappa\ge 0$ and $-2\kappa\le\barkappa\le0$ in
Eq.~(\ref{eq:Helfrich}).

The first term in Eq.~(\ref{eq:Helfrich}) has the standard form of a
surface-tension contribution.  However, inside inclusions, it
represents the elastic energy associated with the in-plane stress induced by
the membrane surface tension.  In the absence of line tension at the
inclusion boundaries, the diagonal components of the equilibrium stress
tensor in the inclusions coincide with the surface tension of the membrane,
which gives $\sigma=\sigma_0$ in Eq.~(\ref{eq:Helfrich}).

The terms with $\kappa$ and $\barkappa$ in Eq.~(\ref{eq:Helfrich}) represent
the bending energy contributions associated with the mean ($\kappa\equiv
\kappa_0+\lambda\kappa_1(\vect{r})$) and Gaussian ($\barkappa\equiv
\barkappa_0+\lambda\barkappa_1(\vect{r})$) curvatures respectively.  While
$\kappa_0$ and $\barkappa_0$ correspond to an unperturbed membrane,
$\kappa_1(\vect{r})$ and $\barkappa_1(\vect{r})$ are position-independent
inside, and vanish outside of the inclusions.

The partition function $\mathcal{Z}$ of an inhomogeneous membrane can
be found as a Boltzmann sum over all membrane configurations
$u(\mathbf{r})$.  With quadratic Eq.~(\ref{eq:Helfrich}), the free
energy, $ \mathcal{F}\equiv -\beta^{-1}\ln
\mathcal{Z}=(2\beta)^{-1}\sum_n\ln (\beta E_n)+\mathrm{const}$, is
expressed in terms of the eigenvalues $E_n$ of the hermitian
``Hamiltonian'' operator $\hat{\mathcal{H}}\equiv
\hat{\mathcal{H}}_{\vect{r}}$ obtained as the second functional
derivative of Eq.~(\ref{eq:Helfrich}) over $u({\bf r})$.  The
corresponding set of orthonormal eigenfunctions $u_n\equiv u_n({\bf
  r})$, $\hat{\mathcal{H}}u_n=E_n u_n$, is complete in the space of
the functions with continuous second derivatives on ${\cal A}$.  We
define the Green's function (GF) of the operator $\hat{\mathcal{H}}$,
\begin{equation}
  \label{eq:gf}
  \hat G\equiv  G(\vect{r},\vect{r}')=\sum_n 
  %% (E_n+0)^{-1}\,
 \frac {u_n(\vect{r}) u_n(\vect{r}')} {E_n},
\end{equation}
which obeys the usual equation
$\hat{\mathcal{H}}_{\vect{r}}G(\vect{r},\vect{r}')=\delta(\vect{r}-\vect{r}')$. 
For a uniform membrane, the term with
$\barkappa$ in the integrand of Eq.~(\ref{eq:Helfrich}) 
is a total derivative; the corresponding energy operator is $\hat
{\mathcal{H}}_0=\kappa_0 
(\nabla^4-\alpha_0^2\nabla^2)$, with $\alpha_0^2\equiv
{\sigma_0/\kappa_0}$.  
The uniform-membrane GF
 is a combination of GFs for the Laplace and the Helmholtz
equations \cite{Evans-Turner-Sens-2003},% 
\begin{equation}
  \label{eq:gf-bare}
  G_0(r)=-\frac{1}{ 2\pi \sigma_0}\bigl[K_0(\alpha_0 r)+\ln (\alpha_0
  r)\bigr],  
  \quad r\equiv|\vect{r}-\vect{r}'|.
\end{equation}

Note that the operator $\hat{\mathcal{H}}\equiv
\hat{\mathcal{H}}_0+\lambda\hat{\mathcal{V}}$, the eigenfunctions $u_n\equiv
u_n^\lambda$, the energies $E_n\equiv E_n^\lambda$, and the free energy
$\mathcal{F}\equiv \mathcal{F}_\lambda$ depend on the coupling parameter
$\lambda$, with $\lambda=0$, corresponding to a non-perturbed (uniform)
membrane.  We use the Hellmann-Feynman theorem
\cite{Hellmann-1937,*Feynman-1938} to write the part of the free energy
associated with the inclusions, $\Delta\mathcal{F}_\lambda\equiv
\mathcal{F}_{\lambda}-\mathcal{F}_0$, as an integral over the parameter $\lambda$,
%%% $=\int_0^\lambda d{\lambda'} \,{\partial
%%%   \mathcal{F}_{\lambda'}\over \partial \lambda'} =(2\beta)^{-1}\sum_n
%%% \int_0^\lambda {d\lambda' \over E_n^{\lambda'}}\,{\partial
%%% E_n^{\lambda'}\over 
%%%   \partial\lambda'}$.  Using the Hellmann-Feynman theorem
%%% \cite{Hellmann-1937,*Feynman-1938}, ${\partial E_n^\lambda\over
%%%   \partial\lambda}=\Bigl\langle u_n^\lambda\Bigl|{\partial
%%%   \hat{\mathcal{H}}\over \partial\lambda}\Bigr| u_n^\lambda\Bigr\rangle
%%% =\langle u_n^\lambda|\hat {\mathcal{V}}| u_n^\lambda\rangle$, we find for
%%% the 
%%% free energy change due to inclusions,
\begin{equation}
\beta\,\Delta
\mathcal{F}_\lambda
%%% =\frac{1}{2}\int_0^{1}d\lambda\sum_n\frac{\langle
%%%  u_n|\hat{\cal V}|u_n\rangle}{E_n}
=\frac{1}{2}\int^\lambda_0 d\lambda'\;\mbox{Tr} (\hat{\cal V} \hat G_{\lambda'}).  
\label{eq:free-energy-gf}
\end{equation}
%%% The matrix elements of the hermitian operator $\hat{\mathcal{V}}$ are given
%%% explicitly by the integral over the inclusions' area,
%%% %%% \begin{eqnarray}
%%% %%%   \label{eq:trace}
%%% %%% \mbox{Tr}(\hat{\mathcal{V}}\hat G)& =&\int d^2\vect{r} \,\Bigl\{
%%% %%%     %\sigma_1(\vect{r}) [\nabla\cdot \nabla'\hat G]+
%%% %%%     \kappa_1(\vect{r}) \,
%%% %%%   \nabla^2 (\nabla')^2 \,\hat G\nonumber\\
%%% %%%   & & %\qquad\qquad \qquad 
%%% %%%   +\,  \barkappa_1(\vect{r})\left [\partial_x^2
%%% %%%   \partial_{y'}^2\,\hat G+\partial_{x'}^2
%%% %%%   \partial_{y}^2\,\hat G
%%% %%%   -2\partial_x\partial_y\partial_{x'}\partial_{y'}\,\hat
%%% G\right]}\Bigr\}_{\vect{r}' 
%%% %%%   =\vect{r},\qquad  
%%% %%% \end{eqnarray}
%%%  \begin{equation}
%%%    \label{eq:matrix-V}
%%%    \langle u,\hat {\mathcal{V}}  v\rangle\equiv
%%%    \!\int\! d^2\mathbf{r}\, \kappa_1\nabla^2 u \nabla^2
%%%    v+\bar\kappa_1(u_{xx}'' 
%%%    v_{yy}''+u_{yy}'' v_{xx}''-2 u_{xy}'' v_{xy}''),
%%% \end{equation}
%%% where the surface tension term is absent since
%%% $\sigma(\mathbf{r})=\sigma_0$. 
For $k>1$ membrane inclusions, we further decompose the
free energy into the part associated with individual inclusions,
$\sum_{l=1}^k\Delta\mathcal{F}_\lambda^{(l)}$, and the Casimir energy
proper, 
\begin{equation}
  \mathcal{F}_{\rm C}\equiv \Delta\mathcal{F}_\lambda-
  \sum\nolimits_{l=1}^k\Delta\mathcal{F}_\lambda^{(l)}.\label{eq:Casimir-proper} 
\end{equation}
To this end we write $\V=\sum_{l=1}^k \V_l$, where the operator
$\V_l$, $l=1,2,\dots, k$, is only non-zero inside the corresponding
inclusion.  Then, if we expand the GF in the integrand of
Eq.~(\ref{eq:free-energy-gf}) in the perturbation series over
$\lambda'$, the $n$-th term is a loop with the unperturbed $\hat G_0$
connecting exactly $n$ operators $\V_l$.  We group together subsequent
terms corresponding to the same inclusion $l$ by introducing the exact
GF for a single inclusion,
\begin{equation}
  \label{eq:exact-gf-l}
  \hat G_\lambda^{(l)}\equiv 
  \hat G_0-\lambda \hat G_0\V_l\hat  G_0+
  \lambda^2 \hat G_0\V_l\hat G_0 \V_l\hat G_0-\cdots. 
\end{equation}
Respectively, the Casimir free energy~(\ref{eq:Casimir-proper}) comprises the
terms with at least two unequal inclusion
indices\cite{Lin-Zandi-Mohideen-Pryadko-long},
\begin{equation}
  \label{eq:free-energy-separated}
  \beta\mathcal{F}_\mathrm{C}=-\!
%  \!\sum_{l_1,...,l_n=0}^k
  \sum_{n>1} 
  \frac{(-\lambda)^{n}}{ 2n}   \sum_{\{ l_i\}}\tr
  \V_{l_1}\hat G_\lambda ^{(l_1)} \V_{l_2}\hat G_\lambda ^{(l_2)}\ldots
   \V_{l_n}\hat G_\lambda ^{(l_n)}, 
\end{equation}
where the $n$-th term involves the summation over $n$ inclusion
indices $1\le l_i\le k$, with the neighboring indices different,
$l_{i+1}\neq l_i$, $l_n\neq l_1$.  Physically, this can be interpreted
as fluctuations' back-and-forth ``hopping'' between the inclusions.
To evaluate the resulting series, we introduce the $k\times k$
operator-valued matrix $\hat\Sigma$ with elements $\Sigma_{ll'}\equiv
(1-\delta_{ll'})\V_l\hat G_\lambda^{(l)}$ (no summation), and write
the inner sum in the $n$\,th term of
Eq.~(\ref{eq:free-energy-separated}) as
\begin{equation}
  \tr \hat\Sigma^n\equiv\sum_{\{l_i\}}\tr \Sigma_{l_1
    l_2}\Sigma_{l_2 l_3}\ldots \Sigma_{l_{n-1}l_n} \Sigma_{l_{n}l_1}, \: n>1,
  \label{eq:sum-of-traces} 
\end{equation}
where neighboring indices are automatically different.
%%% , $l_2\neq l_1$, $l_3\neq l_2$, \ldots, $l_n\neq l_1$. 
Now, the Casimir free energy becomes 
\begin{equation}
  \label{eq:exact-log}
  %%% \Delta\mathcal{F}_\lambda=\sum_l\Delta\mathcal{F}_\lambda^{(l)}+
  %%% \mathcal{F}_\mathrm{C},\;\, 
  \beta\mathcal{F}_\mathrm{C}\equiv
  \frac {1}{2}\tr\, \log
  \bigl(\openone+\lambda\hat\Sigma\bigr).%
\end{equation}
%%% The first term here is the sum of self-energies of individual
%%% inclusions resulting from the $n=1$ term in
%%% Eq.~(\ref{eq:free-energy-separated}).  The second term,
This equation %Eq.~(\ref{eq:exact-log}) 
is our main general result: it is exact, remains
finite in the limit $\lambda\to\infty$, and can be applied to calculate the
Casimir forces between a finite number of compact objects of arbitrary shape
and separation.  The result can also be
recast\cite{Lin-Zandi-Mohideen-Pryadko-long} in a form similar to that of the
scattering matrix approach employed to calculate the electromagnetic (EM)
Casimir interaction\cite{Rahi-2009}.

We now focus on the case of two uniform circular disks embedded in a membrane.
The corresponding matrix $\hat\Sigma$ is $2\times 2$, with no diagonal
elements, and, therefore, the terms with odd powers of $\lambda$ in the
expansion of Eq.~(\ref{eq:exact-log}) disappear.  The matrices $\hat\Sigma^n$,
with even powers $n=2s$, are diagonal, since $\hat\Sigma^2=\mathop{\rm
  diag}(\V_1\hat G_\lambda^{(1)}\V_2\hat G_\lambda^{(2)}, \V_2\hat
G_\lambda^{(2)}\V_1\hat G_\lambda^{(1)})$.  The two matrix elements
give equal contributions to the trace,
%%% Since there are only two inclusions, the $\hat\Sigma$ matrix is $2\times
%%% 2$  
%%% $$
%%% \hat\Sigma=\left  ( 
%%% \begin{array}{cc}
%%% 0                       &  \lambda \Sigma_{12} \\
%%% \lambda\Sigma_{21}   &    0                      
%%%  \end{array}
%%%  \right )
%%%  =
%%%  \left  ( 
%%% \begin{array}{cc}
%%% 0                       &   \lambda \V_1 G_\lambda^{(1)} \\
%%%  \lambda \V_2 G_\lambda^{(2)}  &    0                      
%%%  \end{array}
%%%  \right )
%%% $$
%%% Then the Casimir energy becomes
\begin{eqnarray}
  \beta\,\mathcal{F}_\mathrm{C}
  &=&-\frac{1}{2}\sum_s 
  \frac{\tr[\lambda^2\V_1\hat G_\lambda^{(1)}\V_2\hat G_\lambda^{(2)}]^s}{s}\nonumber\\
  &=&\frac{1}{2}\tr\,\log
  ({\openone-\lambda^2\V_1\hat G_\lambda^{(1)}\V_2\hat G_\lambda^{(2)}}).
  \label{eq:twodisk-full-result} 
\end{eqnarray}

For actual calculations, we construct the exact single-disk GFs $\hat
G_\lambda^{(l)}$, $l=1,2$, as a series in polar coordinates, including
the terms corresponding to both the Laplace ($\propto r^{\pm m}$) and
the Helmholtz [$\propto K_m(\alpha r), I_m(\alpha r)$] equations [cf.\
  Eq.~(\ref{eq:gf-bare})], with the asimuthal quantum number $m<
m_\mathrm{max}$.  This requires four boundary conditions on the
circumference: continuity of the function, normal derivative, as well
as of the following two quantities,
\begin{eqnarray}
  Q_3&\equiv& \sigma \partial_r
  u-\kappa\partial_r\left(\nabla^2u\right)
  +\frac{\bar{\kappa}}{ r}\partial_{r}
  \Bigl(\frac{1}{r}u''_{\theta\theta}\Bigr), 
  \label{eq:bc-3}
  \\
   Q_{4}&\equiv& \kappa \nabla^2u
   +\frac{\bar{\kappa}}{r}\Bigl(\frac{1}{r}u''_{\theta\theta}+\partial_ru\Bigr).
   \label{eq:bc-4}
\end{eqnarray}
Here $u''_{\theta\theta}$ is the second derivative over the polar
angle with respect to the center of the disk.  The same boundary
conditions are used to evaluate the matrix elements of the operators
$\V_l$ in Eq.~(\ref{eq:twodisk-full-result}).  This gives the argument
of the logarithm in Eq.~(\ref{eq:twodisk-full-result}) as a
$2m_\mathrm{max}\times 2m_\mathrm{max}$ matrix (mode doubling corresponds to Helmholtz/Laplace components), which is non-diagonal
since the GFs are
%the two arguments of the GF are
expanded with respect to two different centers.

%%% Multipole expansion around
%%% the inclusions $l$ and $l'$ could be used to express the matrix
%%% elements of $\Sigma_{ll'}$.  

%%% For the numerics, we keep a finite number of azimutal angular
%%% harmonics, $m$.  
At large separations, it is sufficient to keep the
terms up to quadrupole ($m_\mathrm{max}= 2$).  As the distance $R$ between the
disks decreases, higher order multipoles become relevant.  Generally,
when the distance between the edges of the disks, $H\equiv R-2a$, is
small, a large number of multipoles, $m_\mathrm{max}\agt a/H$, is required for
convergence.  In this regime the Casimir energy becomes large, and
independent of the surface tension (see Fig.~\ref{fig:scaling}).

The short separation asymptotic form of the Casimir energy can also be
evaluated within PFA\cite{Derjaguin-1956}.  Here, the interaction
between curved edges is expressed as a sum over infinitesimal straight
line segments approximated as parallel.  We found that the Casimir
energy per unit length for two half-planes is
$\beta\mathcal{F}_\mathrm{C}/L=f/H$, with $f=\pi/24$ in the limit
dominated by the surface tension, and $f\approx 0.46$ in the limit
dominated by the bending energy (it is the latter limit that is
relevant at very small $H$ for finite $\ell_0$).  Then, for 
%%% the special case of 
two hard disks, PFA gives 
\begin{equation}
  \label{eq:pfa}
  \beta \mathcal{F}_{\rm PFA}=-{\pi f}\Bigl[x^{-1/2}+{1\over 2}-{3\over 8}
  x^{1/2}+\mathcal{O}(x)\Bigr], \quad x\equiv {H\over a}.
\end{equation}
We plot the ratio of the Casimir energy, $\mathcal{F}_\mathrm{C}$
[Eq.~(\ref{eq:twodisk-full-result})] calculated for different cutoff $m$ in the regime
dominated by the bending energy, and $\mathcal{F}_{\rm PFA}$
[Eq.~(\ref{eq:pfa})] in the inset of Fig.~\ref{fig:scaling}, and find
that the ratio $\mathcal{F}_\mathrm{C}/\mathcal{F}_\mathrm{PFA}$ approaches
one only at short separations.  As the figure shows the higher order
multipoles are necessary at shorter separations.

%%% importance of including the higher order multipoles is revealed in the
%%% inset of Fig. 1 which shows the ratio of our results to those of PFA
%%% for different values of $m$ at separations $0<H_0/a<0.4$. We need
%%% about $m=100$ partial waves to obtain an accurate results for at
%%% $H_0/a=0.05$.  Higher order multiples are necessary at shorter
%%% separations. At large separations, the ratio goes to zero indicating
%%% that PFA only works at very short separations.

Analytical results for the Casimir free energy~(\ref{eq:twodisk-full-result}) can also
be obtained in the weak coupling regime (small $\lambda$), regardless of the
separation distance $R$, or for any $\lambda$ if $R\gg a$.  These regimes
correspond to keeping the first 
term in the expansion of the logarithm in Eq.~(\ref{eq:twodisk-full-result}).  For the
weak coupling regime (small $\lambda$), we can further simplify the
calculations by replacing $G_\lambda^{(l)}$ with the bare GF, $G_0$.  The full
analytical expressions\cite{Lin-Zandi-Mohideen-Pryadko-long} are too
cumbersome to quote here, and we only present simplified results for three
important parameter ranges.

(\textbf{a}) $\ell_0\gg R$, regime dominated by the bending rigidities of the
membrane.  The Casimir energy~(\ref{eq:twodisk-full-result}) has the asymptotic
form~(\ref{eq:scaling}) with $n=4$ and the coefficient
\begin{eqnarray}
  \label{eq:large-ell-limit}
  A&=&(4B_2^g+A_0^f)  B_2^g,
  \\   \nonumber
  B_2^g&=&{\barkappa_0-\barkappa\over
    4\kappa_0+\barkappa_0-\barkappa},\ \ 
  A_0^f= {4(\kappa-\kappa_0)+2(\barkappa-\barkappa_0)\over 
    2\kappa+\barkappa-\barkappa_0}.
\end{eqnarray}
It is remarkable how the Casimir energy depends on the flexibility of disks in
this regime. For inclusions with finite rigidities,
Eq.~(\ref{eq:large-ell-limit}) is proportional to $\barkappa-\barkappa_0\equiv
\lambda\barkappa_1$, i.e., a discontinuity in $\barkappa$ is required for a
non-zero Casimir force.  The general expression (\ref{eq:large-ell-limit})
reproduces the results obtained previously in two limiting cases.  In the
rigid-disk limit\cite{Goulian-Bruinsma-Pincus-1993,%
  *Goulian-Bruinsma-Pincus-1993B,%
  Golestanian-Goulian-Kardar-PRE-1996,*Park-Lubensky-1996}, where both
$\kappa$ and $-\barkappa$ are infinite, Eq.~(\ref{eq:large-ell-limit}) gives
$A=6$ (horizontal portion of the line shown with red symbols in
Fig.~\ref{fig:scaling}).  In the weak-coupling
limit\cite{Goulian-Bruinsma-Pincus-1993,%
  *Goulian-Bruinsma-Pincus-1993B,Park-Lubensky-1996}, it gives $A=-\lambda^2
\kappa_1 \barkappa_1/2\kappa_0^2$. 

(\textbf{b}) $\ell_0\ll a\ll R$, regime dominated by the surface tension of
the membrane.  We find that the leading-order power law term in $a/R$
(resulting from dipole-like fluctuations around the inclusions) is zero, and
the next-order terms give the Casimir energy~(\ref{eq:twodisk-full-result}) falling
off much faster, $ {\cal F}_\mathrm{C}\propto 1/R^8$, with the coefficient
proportional to $\barkappa_1^2$.    The full result being too bulky,
we only present the strong-coupling limit,
$\beta\Delta\mathcal{F}_\mathrm{C}=-9(a/R)^8$ (in agreement with
Refs.~\onlinecite{Noruzifar-Oettel-2009,Yolcu-Rothstein-Deserno-2010}), and
the leading-order contribution in $\lambda$ at weak-coupling,
\begin{equation}
  \label{eq:power8}
\beta\,\mathcal{F}_\mathrm{C}^{(2)}=
-{36\lambda^2\barkappa_1^2 a^4\over
  \sigma_0^2 R^8}= -{36(\barkappa-\barkappa_0)^2 a^4\over \sigma_0^2 R^8}.
\end{equation}
Note that we also
obtain the same power law in the presence of line tension energy on the
inclusions boundary, in which case Eq.~(\ref{eq:Helfrich}) has
$\sigma\neq\sigma_0$ inside inclusions\cite{Lin-Zandi-Mohideen-Pryadko-long}.
While a power law $\propto 1/R^8$ has been previously obtained
\cite{Noruzifar-Oettel-2009,Yolcu-Rothstein-Deserno-2010} for hard inclusions
in soap films ($\kappa_0=\barkappa_0=0$), we find it remarkable that the
Casimir energy~(\ref{eq:power8}) depends on the difference between the
Gaussian rigidity of inclusions and that of the membrane.

(\textbf{c}) $a\alt\ell_0\alt R$, with both surface tension and bending
rigidities of the membrane relevant.  The full analytical expression
\cite{Lin-Zandi-Mohideen-Pryadko-long} for the Casimir energy between two
disk-like inclusions contains terms decaying like an inverse power of the
distance $R$, and exponentially decaying terms $\propto K_m(\alpha_0R)$.  In
particular, for $\ell_0\ll R$, the terms $\propto K_m(\alpha_0R)$ are
exponentially small.  In this case the Casimir interaction energy scales
$\propto 1/R^8$, with the coefficient which is a complicated function of
parameters, especially in the region $\ell_0\sim a$.  The exponentially small
terms become relevant when $\ell_0\sim R$, where Casimir energy crosses over
to the small-$\sigma$ regime (\textbf{a}) with ${\cal F}_\mathrm{C}\propto
1/R^4$.
This crossover can be seen in Fig.~\ref{fig:scaling}.  A representative case
corresponds to $\ell_0=10a$, where $A\equiv -(R/a)^4\mathcal{F}_\mathrm{C}$ is
nearly constant for $R\lesssim \ell_0$, is strongly reduced for larger $R$,
and eventually crosses over to $\propto 1/R^4$
(${\mathcal{F}}_\mathrm{C}\propto 1/R^8$) for $R\gg \ell_0$.  At smaller $R$,
the same asymptotic power law is also seen, e.g., for $\ell_0/a=1$.  

Note that the distance dependence of the Casimir energy is the same, $\propto
1/R^8$, as long as $\ell_0\ll R$, which includes regimes (\textbf{b}) and
(\textbf{c}).  In the regime (\textbf{b}), dominated by the surface tension,
this power law can be obtained by treating the inclusions as point-like
objects in the effective field theory
(EFT)\cite{Yolcu-Rothstein-Deserno-2010}.  For inclusions that are free to
tilt with the membrane, the expansion starts with the quadrupole
terms\cite{Yolcu-Rothstein-Deserno-2010}.  In the regime (\textbf{c}) the
higher-order multipole terms in the EFT expansion diverge as increasing powers
of $\ell_0/a\gg1$.  However, the contributions to the Casimir energy coming
from higher-order multipole terms also get suppressed as increasing powers of
$1/R$.  As a result, the leading-order quadrupole terms dominate, which again
gives $\mathcal{F}_\mathrm{C}\propto 1/R^8$ for $\ell_0\ll R$.  For
$\ell_0\sim R$, where we recover the exponentially small terms $\propto
K_m(R/\ell_0)$, all multipoles contribute equally and the EFT approach cannot
be used.

In conclusion, we have developed an exact method for computing the Casimir
energy between elastic inclusions of arbitrary shapes embedded in a biological
membrane under tension, characterized by the surface tension $\sigma_0$ and
bending and Gaussian rigidities, $\kappa_0$ and $\barkappa_0$. The method
allows to calculate the Casimir forces in all ranges of parameters and for all
separations.  The Casimir energies are fully characterized by the objects'
``scattering'' matrices, which encode the shapes and mechanical properties.
In particular, for two elastic disks, the Casimir energy scales as $\propto
1/R^4$ for $R\lesssim \ell_0$, and crosses over to $\propto 1/R^8$ for $R\agt
\ell_0$.  At short distances, the Casimir energy is large; for hard disks our
findings agree with the corresponding PFA results,
$\mathcal{F}_\mathrm{C}\propto H^{-1/2}$.  One
interesting result is that the Casimir energy is strongly suppressed for
inclusions whose Gaussian rigidity $\barkappa$ equals that of the membrane.

We are pleased to acknowledge important discussions with M.~Kardar.
This research was supported in part by NSF grants DMR-0645668 (RZ),
CCF-0622242 (LP), PHY-0970161 (UM), DOE DE-FG02-04ER46131 (UM), and DARPA
N66001-091-2069 (UM \& RZ).

%\bibliography{casimir,membrane,lpp,add}
%merlin.mbs 2010-03-15 4.21a (PWD, AO, DPC)

\end{document}